\documentclass[pre, showpacs, aps]{revtex4}

\usepackage{graphicx}

\newcommand{\pd}[2]{\frac{\partial #1}{\partial #2}}
\newcommand{\dd}[2]{\frac{d #1}{d #2}}

\newcommand{\ts}{{t^*}}
\newcommand{\ksr}[1]{{{k_*^{(R)}}(#1)}}
\newcommand{\kep}[1]{{k_\epsilon(#1)}}

\begin{document}

\author{Colm Connaughton}
\affiliation{ Laboratoire de Physique Statistique de l'Ecole
Normale Sup\'erieur, associ\'e au CNRS, 24 Rue Lhomond, 75231
Paris Cedex 05, France}
\author{Sergey Nazarenko}
\affiliation{Mathematics Institute, University of Warwick,
Coventry CV4 7AL, United Kingdom}
\date{\today}
\title{A model differential equation for turbulence}

\begin{abstract}
A phenomenological turbulence model in which the energy spectrum
obeys a nonlinear diffusion equation is presented. This equation
respects the scaling properties of the original Navier-Stokes
equations and it has the Kolmogorov $-5/3$ cascade and the
thermodynamic equilibrium spectra as exact steady state solutions.
The general steady state in this model contains a nonlinear
mixture of the constant-flux and thermodynamic components. Such
``warm cascade'' solutions describe the bottleneck phenomenon of
spectrum stagnation near the dissipative scale. Self-similar
solutions describing a finite-time formation of steady cascades
are analysed and found to exhibit nontrivial scaling behaviour.
\end{abstract}


\maketitle

\section{Model Equation}
In this letter, we present a model equation for the energy
spectrum of isotropic homogeneous turbulence,
\begin{equation}
E_t = {1 \over 8} (k^{11/2} E^{1/2} (E/k^2)_k)_k +f - \nu k^2 E,
\label{fp}
\end{equation}
where $t$ is time, $k$ is the absolute value of the wavenumber,
$\nu$ is the kinematic viscosity coefficient, $f$ is an external
forcing and the energy
spectrum $E(k,t)$ is normalised so that the kinetic energy density
is $\int E \, dk$.
The third term on the RHS of this equation is obvious and describes
the linear process of viscous dissipation.
The first term contains a nontrivial model of the nonlinear processes
in turbulence which rests on three basic properties\footnote
{The parts 1 and 2 were inspired by paper by Kulsrud and Anderson
\cite{kulsrud}. However, they used a 1st order
in $k$ equation which does not allow thermodynamic equilibria and
which fails to describe spreading of the spectrum toward lower $k$.}:

1. The total energy density  $\int E \, dk$ is conserved by the
inviscid dynamics. The characteristic time of the spectral energy
redistribution for {\em local} interaction of turbulent scales is
of order of the vortex turnover time, $1/\sqrt{k^3 E}$.

2. The steady state in forced turbulence corresponds to a constant
energy cascade though the inertial range of scales which is
described by the Kolmogorov spectrum,
\begin{equation}
E = C \, P^{2/3} \, k^{-5/3},
\label{kolm}
\end{equation}
where $P$ is the energy flux (constant in $t$ and $k$) and $C$ is
the Kolmogorov constant. Experimental measurements \cite{sreeni}
give $C=1.6 \pm 0.17$. As we will see below, equation (\ref{fp})
has an exact solution of form (\ref{kolm}) with $C = (24/11)^{2/3}
\approx 1.68$ ($C$ can be changed by tuning the constant factor in
the first term).

3. When the wave-number range is truncated to a finite value and
both forcing and dissipation are absent,  turbulence reaches a
thermodynamic equilibrium state characterized by equipartition of
energy over the wave-number space \cite{kraichnan}.
 In terms of the one-dimensional energy spectrum this means
$E \propto k^2$ which is obviously a steady state solution of the
equation (\ref{fp}) for $f= \nu =0$. Mathematical simplicity and
respect to the above  basic properties of Navier-Stokes turbulence
make model (\ref{fp})  useful for practical and numerical
applications. We will now analyze solutions of (\ref{fp}) in
greater detail to find other properties that are predicted by this
model.

\section{Stationary Solutions}

Let us consider steady-state spectra in the inertial range. For
$f= \nu =0$, we have the following general time-independent
solution,
\begin{equation}
E = C \, k^2 \, (P k^{-11/2} + Q)^{2/3},
\label{steady}
\end{equation}
where $C=(24/11)^{2/3} \approx 1.68$ and $P$ and $Q$ are arbitrary
constants. For $Q=0$, this gives the pure Kolmogorov cascade
solution (\ref{kolm}), whereas for $P=0$ this is a pure
thermodynamic equilibrium. For the general solution, both the
constant  flux of energy $P =  - {1 \over 8} k^{11/2} E^{1/2}
(E/k^2)_k \ne 0$ and a thermodynamic part ($Q \ne 0$) are present;
they appear as an interesting nonlinear combination and not just
as a linear superposition because equation  (\ref{fp}) is
nonlinear. Thus, one can refer to solution (\ref{steady}) with
finite $P$ and $Q$ as a {\em warm cascade} to distinguish it from
the pure Kolmogorov solution which could be viewed as a {\em cold
cascade}.

\begin{figure}
\begin{center}
\includegraphics[width=4.0in]{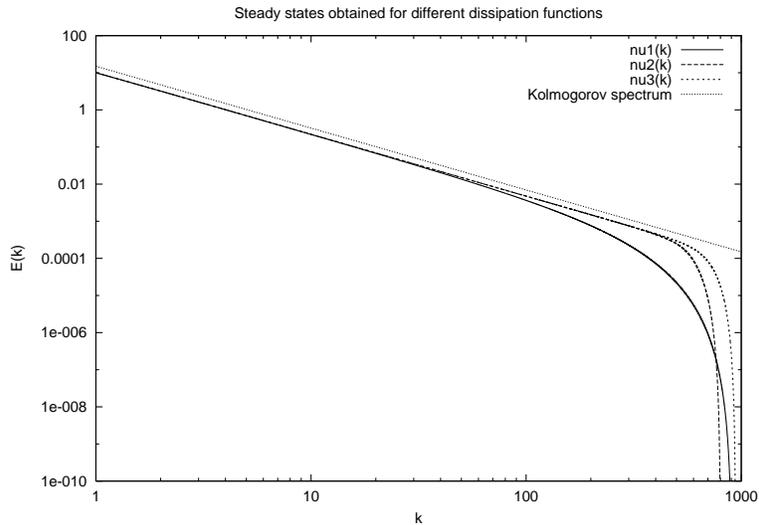}
\end{center}
\caption{\label{fig-KCascades}Numerically computed steady states
for several choices of dissipation function: $\nu_1(k) =
(500-k)^2$, $\nu_2(k) = 4.0\times 10^{-6} (500-k)^4$, $\nu_3(k) =
1.0\times 10^{-2}k^2$. The Kolmogorov spectrum is also shown for
comparison but shifted slightly for clarity.  }
\end{figure}

Let us suppose that turbulence is produced near some scale $k=k_0$
so that $f(k) >0$ only  in a finite range in the vicinity of
$k_0$. Suppose that $f(k) =0$ to the left of this range (at large
scales) and in a large inertial range to the right which ends at a
very high $k \sim k_d$ where viscosity $\nu$ or some other
dissipation mechanisms ($f(k) < 0$) become important. Then,
up-scale of the forcing there will be a pure thermodynamic
solution with $P=0$ and $Q \ne 0$ because there is no dissipation
or forcing assumed to be present near $k=0$ to absorb or generate
a finite energy flux. In the inertial range down-scale of the
forcing there will be a constant flux cascade solution. This
solution typically takes the form of a pure Kolmogorov (cold)
cascade and extends down to the dissipation range where the energy
flux is absorbed. Typically, the solution only penetrates a finite
distance into the dissipation range and adapts itself until it
provides sufficient dissipation to absorb the supplied flux. In
the presence of dissipation, the model does not develop structure
at arbitrarily high $k$ as it would in the inviscid case. The
qualitative features of the steady state are independent of the
detailed form chosen for the dissipation. Figure
\ref{fig-KCascades} shows the steady state solutions obtained
numerically for several different choices of the dissipation.

\begin{figure}
\begin{center}
\includegraphics[width=4.0in,]{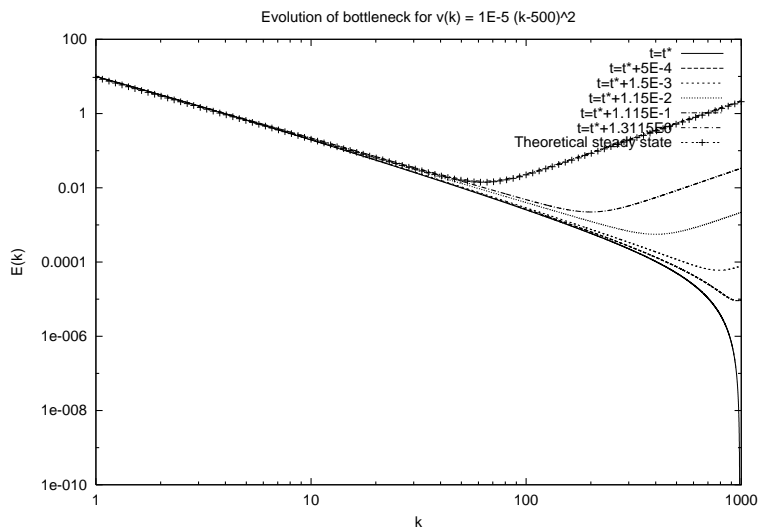}
\end{center}
\caption{\label{fig-bottleneck}Numerical evolution of a bottleneck
for dissipation function $\nu(k) = 1.0\times 10^{-5}(k-500)^2$.
The resulting steady state is well described by the solution
(\ref{steady}).}
\end{figure}

However, if the dissipation is not sufficiently strong, the
solution can penetrate far enough into the dissipation range to
reach the maximal wave-number which necessarily exists in any
numerical solution. If one imposes a zero flux condition at the
right end of the computational interval, the energy flux is
reflected from the maximal wave-number leading to greater values
of $E$ in the dissipative range. Such a cascade stagnation acts to
enhance the dissipation rate and thereby to adjust it to the
energy flux to be absorbed. This phenomenon is common in numerical
simulations of turbulence and is usually called the bottleneck
phenomenon \cite{falkovich}. Figure \ref{fig-bottleneck} shows a
numerically obtained steady state for the dissipation function,
\begin{eqnarray*}
\nu(k)&=& \nu_0 (k-k_D)^2\ k>k_D\\
      & & 0\ k<k_D
\end{eqnarray*}
with $\nu_0=1.0\times 10^{-5}$ and $k_D=500$. The bottleneck
phenomenon is clearly seen as an energy ``pile up'' over the cold
cascade solution near the dissipative scale. In our model, the
bottleneck phenomenon is described by the {\em warm} cascade
solutions; in particular the theoretical curve in figure
\ref{fig-bottleneck} is computed by taking $P\approx 14.5$, $Q
\approx 1.5\times 10^{-9}$ in equation (\ref{steady}). The
relative importance of the ``thermal'' effects with respect to the
cascade grows as one moves from larger to smaller scales; in
extreme cases the spectrum can be nearly pure Kolmogorov near the
forcing range and almost purely thermodynamic near the dissipative
scale.

Because of the nonlinearity, solutions for given forms of forcing and dissipation
are usually hard to find analytically and one has to use numerics.
However, a lot of insight about the qualitative behavior of the system can
be gained from considering stationary solutions (\ref{steady}) in an inertial range
$k_1 < k <k_2$ and fixing the spectrum at its boundaries, $E(k_1) = E_1,
E(k_2) = E_2$. This kind of the boundary conditions roughly models the
forcing and the dissipation effects outside of the inertial range.
This gives
\begin{eqnarray}
 P= [(E_2/Ck_2^2)^{3/2} - (E_1/Ck_1^2)^{3/2}]/(k_2^{-11/2} - k_1^{-11/2})
\\
Q=   [k_2^{5/2} (E_2/C)^{3/2} - k_1^{5/2} (E_1/C)^{3/2}]/(k_2^{11/2} - k_1^{11/2})
\end{eqnarray}
Thus, the sign of  $P$ is opposite to the sign of $(E_2/E_1 -
k_2^2/k_1^2)$ and can be either positive or negative depending of
the spectrum steepness with thermodynamic $k^2$ solution been a
borderline case for which $P=0$. Constant $Q$ can also be either
positive or negative with Kolmogorov $-5/3$  been the borderline
slope. It is convenient to think of the $Q <0$ solutions as
negative temperature states.

\section{Non-stationary Solutions}
So far we concentrated on the stationary solutions but how do
these solutions form? We consider from now on the inviscid case.
The Kolmogorov $-5/3$ spectrum is of a {\em finite capacity} type
in that it contains only a finite amount of energy at the high $k$
end. Let us take an initial condition which is compactly supported
and force the system by imposing a constant flux boundary
condition across the left end of the computational interval. Owing
to the finite capacity of the Kolmogorov solution, an infinitely
remote dissipative scale must be reached in a finite time.  The
solution has a nonlinear front at $k=k_*(t)$ and this front
accelerates explosively, reaching $k=\infty$ at a finite singular
time which we shall denote by $t_*$. We can equally consider the
decaying case where an initial distribution of energy, compactly
supported at large scales, is allowed to spread under the action
of the nonlinearity without any external forcing. Provided there
is sufficient energy in the initial condition, the right front
still reaches $k=\infty$ within a finite time but in addition a
second front propagates to the left spreading the spectrum to
large scales. This second front does not exhibit singular
behaviour in finite time. Hence the decaying case is broadly
similar to the forced case as far as large $k$ behaviour is
concerned since the initial concentration of energy at large
scales acts as an effective forcing for the right front.

It is well known (see for example \cite{lacey}) that the solutions
of nonlinear diffusion equations with compactly supported initial
data often have the property of remaining compactly supported
during the time evolution. This turns out to be the case here. The
left and right nonlinear fronts actually correspond to the left
and right extrema of the support of the solution. Since these
points must be determined as part of the solution, we are, in
principle, required to solve a moving boundary problem with two
free boundaries. In order that the problem be well-posed, we
require an additional {\em moving boundary condition} on each
interface\cite{ockendonsBook}. For the problem under
consideration, the appropriate condition for the right front is
\begin{equation}
\label{eq-MBC} \dd{k_*^{(R)}}{t} = \lim_{k\to k_*^{(R)}}
k^\frac{11}{2} E^{-\frac{1}{2}}\pd{}{k}\left(k^{-2}E\right).
\end{equation}
A similar condition holds for the left front.

\begin{figure}
\begin{center}
\includegraphics[width=4.0in,]{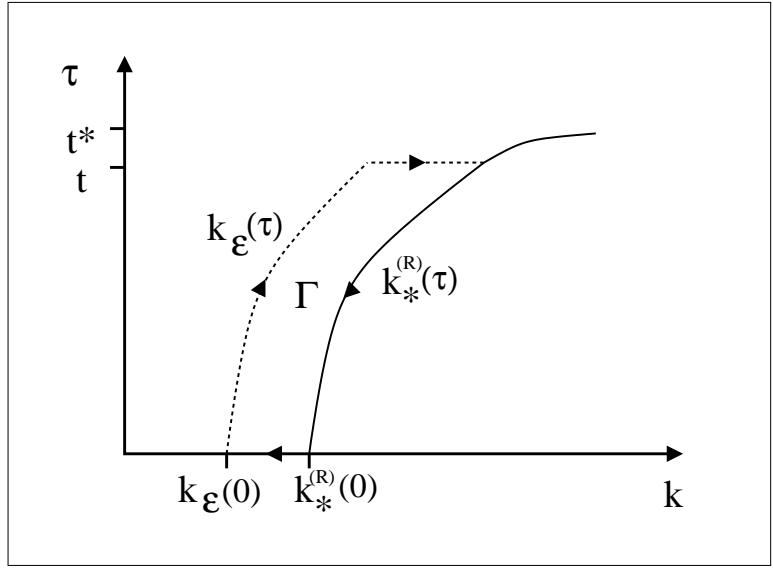}
\end{center}
\caption{\label{fig-MBC}Integration contour leading to the moving
boundary condition, (\ref{eq-MBC}).}
\end{figure}

This formula can be obtained as follows. Consider the time
interval $\left[0,t\right]$ with $t<\ts$. For $0<\tau<t$ the right
interface is given by the curve $k^{(R)}_*(\tau)$ as shown in
figure \ref{fig-MBC}. Taking, $\epsilon <<1$ and assuming that the
solution goes continuously to zero at the right interface we
define the curve $k_\epsilon(\tau)$ by the condition
$E(k_\epsilon(\tau),\tau)=\epsilon$ for $0<\tau<t$. We now
integrate around the contour, $\Gamma$, in $(k,t)$ plane as shown
in the figure,
\begin{equation}
\oint_\Gamma\left(\pd{E}{T}+\pd{P}{k}\right)\,dk\,d\tau =0.
\end{equation}
We obtain
\begin{equation}
\oint_\Gamma\left(E\,dk + P\,d\tau\right) =0.
\end{equation}
Proceed as follows
\begin{equation}
\oint_\Gamma E\, \left(dk +\tilde{P}\,d\tau \right) =0,
\end{equation}
where
\begin{equation}
\label{eq-Ptilde} \tilde{P} =-k^\frac{11}{2}
E^{-\frac{1}{2}}\pd{}{k}\left(k^{-2}E\right)
\end{equation}
\begin{equation}
 \Rightarrow \int_\kep{t}^\ksr{t} E\,dk +\int_\ksr{\tau} E
\left(dk + \tilde{P}\,d\tau\right) + \int_\ksr{0}^\kep{0} E\,dk
+\int_\kep{\tau} E \left(dk + \tilde{P}\,d\tau\right) =0.
\end{equation}
Now $E=0$ on $\ksr{\tau}$ and $E=\epsilon$ on $\kep{\tau}$. Using
these facts we obtain
\begin{equation}
\kep{t} - \kep{0} + \int_{k_\epsilon}\tilde{P}\,d\tau =
\int_\kep{0}^\ksr{0}\frac{E}{\epsilon}\,dk -
\int_\kep{t}^\ksr{t}\frac{E}{\epsilon}\,dk.
\end{equation}
We now take $\epsilon\to 0$ so that $\kep{\tau}\to\ksr{\tau}$. The
integrands on the RHS are bounded by 1 and therefore give no
contribution. We are left with
\begin{equation}
\ksr{t} = \ksr{0} - \int_0^t\tilde{P}(\ksr{\tau},\tau)\,d\tau,
\end{equation}
which in conjunction with (\ref{eq-Ptilde}) yields (\ref{eq-MBC})
upon differentiation.

Let us now look for a self similar solution of equation (\ref{fp})
taking the following form
\begin{equation}
E = (t_* - t)^a F(\eta); \;\;\;\;\; \eta = k /k_*, \;\;\; k_* = c (t_* - t)^b,
\label{self}
\end{equation}
where $a, b $ and $c$ are constants. Clearly, $b$ must be negative
since we require that $k_*\to\infty$ as $t \to t_*$. Substituting
(\ref{self}) into (\ref{fp}) with $f=\nu=0$ we find that the $t$
dependence drops out of the equation if $ a=-2-3b. $ We then have
the following equation for $F$,
\begin{equation}
(3b+2) F  + b \eta F' = {C^{3/2} \over 8} (\eta^{11/2} F^{1/2} (F/k^2)')',
\label{selfeqn}
\end{equation}
where prime means differentiation with respect to $\eta$. Equation
(\ref{selfeqn}) defines a one-parameter family of self-similar
solutions. Note that substitution of the form (\ref{self}) into
the moving boundary condition, (\ref{eq-MBC}) yields the same
similarity relations. Thus the moving boundary condition is
consistent with the similarity ansatz but does not provide any
additional constraints. In particular it does not tell us which
member of the 1 parameter family of solutions is selected by the
PDE.

The solution near the front tip can be found by expanding $F$ in
series with respect to small $(1-\eta)$; in the leading order we
have
\begin{equation}
F={16 b^2 \over C^3} (1-\eta)^2 \label{tip}
\end{equation}
which gives for the spectrum
\begin{equation}
E={16 b^2 \over {k_*}^3 (t_*-t)^2 } \left(1-{k \over k_*}
\right)^2. \label{tipE}
\end{equation}
Notice that the quadratic decay of the solution as one approaches
the tip is exactly the decay required in order to ensure that the
speed of the tip as given by (\ref{eq-MBC}) is finite and
non-zero.

We are interested in solutions which behave like a power law far
behind the front. That is $E \sim k^{-x}$ as $k\to 0$. The
relations (\ref{self}) then imply that $x=-a/b$. The pure
Kolmogorov spectrum, $x=5/3$, therefore requires $b=-3/2$,
corresponding to what one might consider to be normal scaling in
the wake of the front.

\begin{figure}
\begin{center}
\includegraphics[width=4.0in]{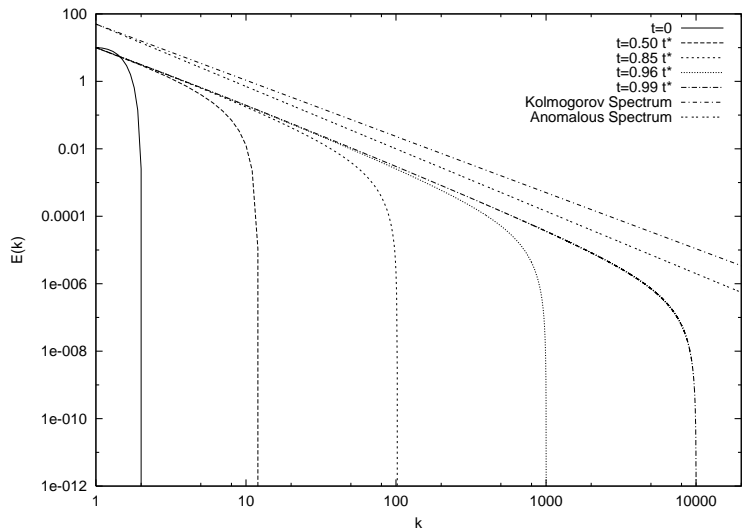}
\end{center}
\caption{\label{fig-front}Forced time-dependent solutions
beginning from compact initial data showing development of
self-similar front with power law wake. }
\end{figure}

\begin{figure}
\begin{center}
\includegraphics[width=4.0in]{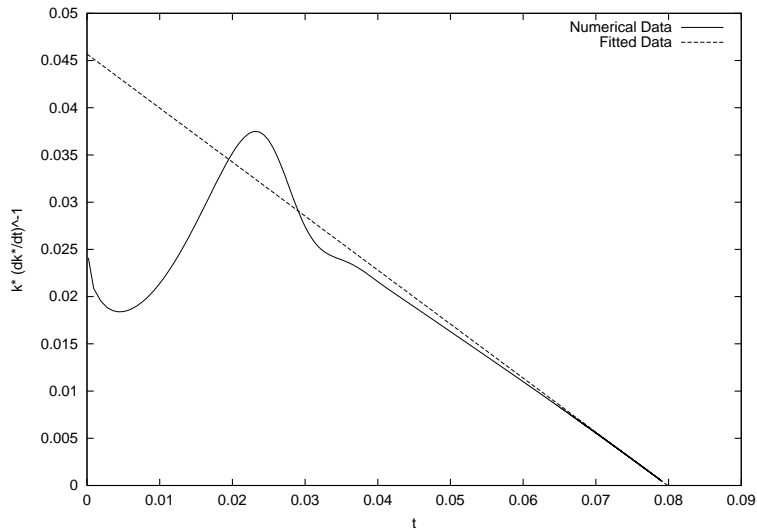}
\end{center}
\caption{\label{fig-profile}Calculation of the asymptotic scaling
properties of the self-similar solution.}
\end{figure}

We performed numerical simulations of the forced solutions of
equation (\ref{fp}) with compact initial data and constant input
flux on the left.  We use the numerical solution to check the
development of a self-similar front with tip of form (\ref{tip})
and to determine which value of $b$ is selected. The results are
shown in figure figure \ref{fig-front}. The scaling parameter, $b$
and the singular time, $t_*$, are most conveniently extracted from
the relation
\begin{equation}
\label{eq-myRelation} k_*\left(\dd{k_*}{t}\right)^{-1} =
-\frac{1}{b}\, (t_* -t),
\end{equation}
which allows one to calculate $b$ and $t_*$ from a linear fit of
the data near $t=t_*$ as shown in figure \ref{fig-profile}. We
find that $t_*=0.0799$ and $b=-1.748$ which corresponds a
significantly steeper than Kolmogorov slope, $x=1.856$. The
singular time, $t_*$ depends on the choice  of initial conditions
but the anomalous scaling exponent does not. In particular we
verified that the same value of $b$ is obtained for the decaying
case. Such anomalous scaling behavior whereby the exponent of the
solution in the wake of the nonlinear front is larger than the
Kolmogorov value has been observed before. Examples include MHD
wave turbulence \cite{alfv} and optical turbulence\cite{nonstat}.
We expect that this anomaly is a general property of finite
capacity systems rather than a property of our model and therefore
may also be present in the Navier-Stokes equations.

\section{Origin of the Transient Spectrum}
For the model (\ref{fp}) the origin of the anomaly can be traced
to the question of existence of a solution of the similarity
equation (\ref{selfeqn}) which has the correct behaviour {\em
both} for the wake and the tip.   Written in terms of $x$ rather
than $b$ and re-scaled to get rid of constants, the similarity
equation is
\begin{equation}
\label{eq-ss2} \frac{2}{x-3}\left(\eta\dd{F}{\eta} + xF\right) =
\dd{}{\eta}\left(\eta^\frac{11}{2}\sqrt{F}\dd{}{\eta}\left(\eta^{-2}F\right)\right).
\end{equation}
We require that this equation have a solution which behaves as
$\eta^{-x}$ as $\eta\to 0$ and also satisfies the boundary
condition, (\ref{eq-MBC}), at the front tip, $\eta\to 1$. Such a
solution is not typical and actually exists only for one value of
$x$. In particular, such a solution does not exist for $x=5/3$,
the Kolmogorov value. The structure of the problem can be studied
by introducing a new independent variable, $s=\log \eta$ and a
pair of dependent variables, $f(s)$, $g(s)$ defined by
\begin{equation}
F = 25\,\eta^{-3}f^2,\ \ \  \dd{F}{\eta} =
\frac{25}{3}\,\eta^{-4}f g.
\end{equation}
The purpose of this transformation is to ``autonomise'' the
equation\footnote{Simpler autonomising transformations can be
certainly be found. The form of the transformation has been chosen
in order to make more convenient the positions of the equilibrium
points of the resulting dynamical system.}. That is, we remove the
explicit dependence on $\eta$ from the equation. When this is
done, equation (\ref{eq-ss2}) is equivalent to the following
autonomous first order system:
\begin{eqnarray}
\label{eq-system}\dd{f}{s}&=& \frac{3}{2}\left(f+g\right)\\
\nonumber f\dd{g}{s}&=&\frac{1}{3}\left(5 f^2+6 f g - 9 g^2
+\frac{10}{x-3}\left(3 f + x g\right)\right),
\end{eqnarray}
and methods of phase plane analysis can be applied. The associated
dynamical system has three equilibria P1 $= (0,0)$, P2 $= (0,
10/3(x-3))$ and P3 $= (1,-1)$. Note that P1 and P2 are singular
points of the original equations, (\ref{eq-system}).

Let us now attempt to identify the trajectory in the $(f,g)$ plane
describing the physical solution.  We note that the required
trajectory must remain in the quadrant $f>0$, $g<0$. The wake is
at $\eta\to 0$, or $s\to \-\infty$. In the wake, $F\sim\eta^{-x}$
and $F^\prime \sim -x\eta^{-x-1}$. It follows that $f\sim
\eta^{3-x}$ and $g\sim -x\eta^{3-x}$.  The value of $x$ is less
than 3 so both $f$ and $g$ go to zero as $\eta\to 0$. The wake is
therefore at $(0,0$ and should be reached as $s\to -\infty$.

Now consider the tip. We know that the front tip is at $\eta=1$,
or $s=0$.Both $F$ and $F^\prime$ go to zero as $\eta\to 1$ but it
follows from (\ref{eq-MBC}) that the product $F^{-1/2}F^\prime$
remains finite. The front tip therefore lies at some finite point
on the negative $g$ axis. In fact the tip lies at P2. To show
this, we show that the moving boundary condition is satisfied
there. Translating (\ref{eq-MBC}) into the self-similar variables
we obtain
\begin{equation}
-b \tau^{b-1} = - \tau^{\frac{5}{2}b+\frac{1}{2} a}\,
\lim_{\eta\to 1}\ \eta^\frac{7}{2}\frac{1}{\sqrt{F}} \dd{F}{\eta}.
\end{equation}
Balancing the $\tau$ dependence gives us back the self-similarity
condition, $a=-2-3b$. We now carefully do the various changes of
variables and rescalings to obtain
\begin{equation}
b=\lim_{s\to 0} \frac{3}{5} h.
\end{equation}
Noting the similarity relations (\ref{self}), we see that this
condition is satisfied identically at the point $P_2$.

\begin{figure}
\begin{center}
\includegraphics[width=4.0in,]{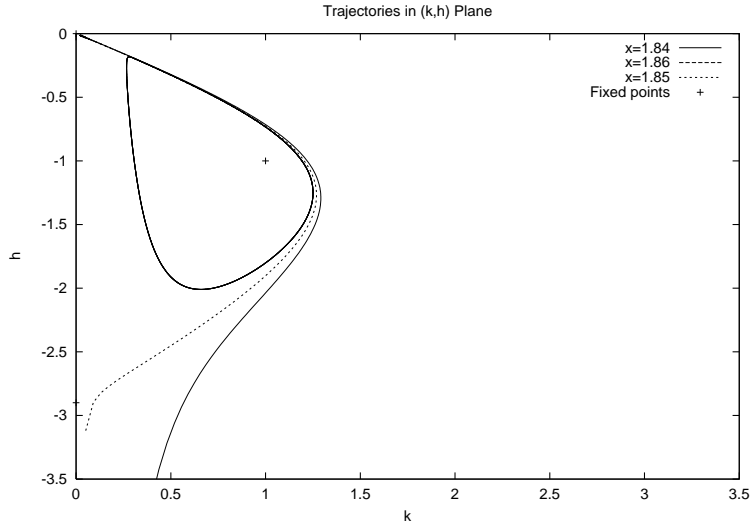}
\end{center}
\caption{\label{fig-phasePlane}Trajectories in the $(k,h)$ plane.}
\end{figure}

\begin{figure}
\begin{center}
\includegraphics[width=4.0in,]{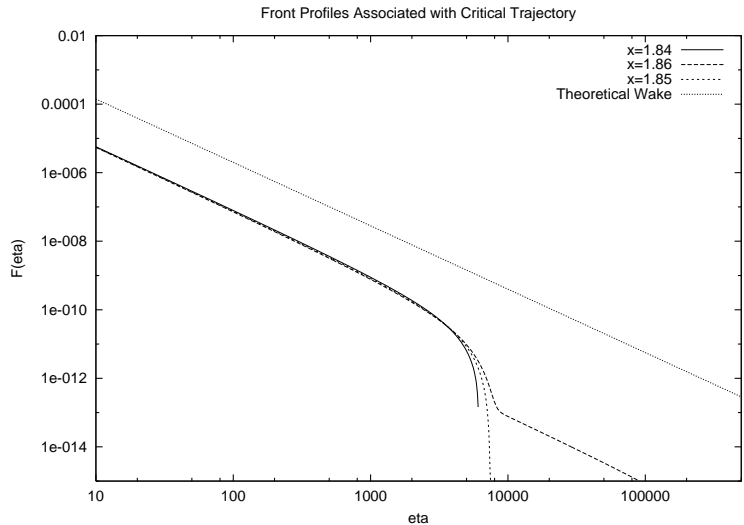}
\end{center}
\caption{\label{fig-profiles}Corresponding front profiles.}
\end{figure}

The critical value of $x$ occurs when  the unstable manifold of
$P_1$ intersects the stable manifold of $P_2$ thus forming a
connection between the two singular points of the original
equations. This is then the only trajectory which can satisfy the
required conditions to describe both the tip and the wake. A
numerical approximation to the unstable manifold of the origin as
$x$ is varied through the critical value is shown in figure
\ref{fig-phasePlane}. The corresponding front profiles, converted
back into the self-similar variables are shown in figure
\ref{fig-profiles}.

 In practice, any model should include dissipation so that
this self-similar solution above will be valid only until the
front tip meets the dissipation scale. After this, the transient
slope gets replaced in the inertial range by the stationary
cascade solution, with or without bottleneck depending on the
dissipation, as discussed above.

In summary, we presented a simple model in which basic properties of the Navier-Stokes
turbulence are built-in: dimensionality and scaling, Kolmogorov and thermodynamic spectra.
This model allows to obtain new predictions about the ``warm'' cascade states which are a
nonlinear mixture of the cascade and thermodynamic solutions and which describe the
bottleneck phenomenon. The model also allowed us to study the self-similar dynamics
of the finite-time formation of the steady cascade states.

We thank Ildar Gabitov and Alan Newell for helpful discussions.

\bibliography{main}

\end{document}